\documentclass[aps,prd,superscriptaddress,showpacs,nofootinbib,floatfix,twocolumn]{revtex4}
\usepackage{graphicx}	% Include figure files

\begin{document}


\title{Inclusive cross section and double helicity asymmetry \\
for $\pi^{0}$ production in $p+p$ collisions at $\sqrt{s} = 200$ GeV:  \\
Implications for the polarized gluon distribution in the proton}

\newcommand{\abilene}{Abilene Christian University, Abilene, TX 79699, U.S.}
\newcommand{\banaras}{Department of Physics, Banaras Hindu University, Varanasi 221005, India}
\newcommand{\bnl}{Brookhaven National Laboratory, Upton, NY 11973-5000, U.S.}
\newcommand{\caucr}{University of California - Riverside, Riverside, CA 92521, U.S.}
\newcommand{\charlesczech}{Charles University, Ovocn\'{y} trh 5, Praha 1, 116 36, Prague, Czech Republic}
\newcommand{\ciae}{China Institute of Atomic Energy (CIAE), Beijing, People's Republic of China}
\newcommand{\cns}{Center for Nuclear Study, Graduate School of Science, University of Tokyo, 7-3-1 Hongo, Bunkyo, Tokyo 113-0033, Japan}
\newcommand{\colorado}{University of Colorado, Boulder, CO 80309, U.S.}
\newcommand{\columbia}{Columbia University, New York, NY 10027 and Nevis Laboratories, Irvington, NY 10533, U.S.}
\newcommand{\czechtech}{Czech Technical University, Zikova 4, 166 36 Prague 6, Czech Republic}
\newcommand{\dapnia}{Dapnia, CEA Saclay, F-91191, Gif-sur-Yvette, France}
\newcommand{\debrecen}{Debrecen University, H-4010 Debrecen, Egyetem t{\'e}r 1, Hungary}
\newcommand{\elte}{ELTE, E{\"o}tv{\"o}s Lor{\'a}nd University, H - 1117 Budapest, P{\'a}zm{\'a}ny P. s. 1/A, Hungary}
\newcommand{\fit}{Florida Institute of Technology, Melbourne, FL 32901, U.S.}
\newcommand{\fsu}{Florida State University, Tallahassee, FL 32306, U.S.}
\newcommand{\gsu}{Georgia State University, Atlanta, GA 30303, U.S.}
\newcommand{\hiroshima}{Hiroshima University, Kagamiyama, Higashi-Hiroshima 739-8526, Japan}
\newcommand{\ihepprot}{IHEP Protvino, State Research Center of Russian Federation, Institute for High Energy Physics, Protvino, 142281, Russia}
\newcommand{\illuiuc}{University of Illinois at Urbana-Champaign, Urbana, IL 61801, U.S.}
\newcommand{\instpasczech}{Institute of Physics, Academy of Sciences of the Czech Republic, Na Slovance 2, 182 21 Prague 8, Czech Republic}
\newcommand{\isu}{Iowa State University, Ames, IA 50011, U.S.}
\newcommand{\jinrdubna}{Joint Institute for Nuclear Research, 141980 Dubna, Moscow Region, Russia}
\newcommand{\kek}{KEK, High Energy Accelerator Research Organization, Tsukuba, Ibaraki 305-0801, Japan}
\newcommand{\kfki}{KFKI Research Institute for Particle and Nuclear Physics of the Hungarian Academy of Sciences (MTA KFKI RMKI), H-1525 Budapest 114, POBox 49, Budapest, Hungary}
\newcommand{\korea}{Korea University, Seoul, 136-701, Korea}
\newcommand{\kurchatov}{Russian Research Center ``Kurchatov Institute", Moscow, Russia}
\newcommand{\kyoto}{Kyoto University, Kyoto 606-8502, Japan}
\newcommand{\labllr}{Laboratoire Leprince-Ringuet, Ecole Polytechnique, CNRS-IN2P3, Route de Saclay, F-91128, Palaiseau, France}
\newcommand{\lawllnl}{Lawrence Livermore National Laboratory, Livermore, CA 94550, U.S.}
\newcommand{\losalamos}{Los Alamos National Laboratory, Los Alamos, NM 87545, U.S.}
\newcommand{\lpc}{LPC, Universit{\'e} Blaise Pascal, CNRS-IN2P3, Clermont-Fd, 63177 Aubiere Cedex, France}
\newcommand{\lund}{Department of Physics, Lund University, Box 118, SE-221 00 Lund, Sweden}
\newcommand{\muenster}{Institut f\"ur Kernphysik, University of Muenster, D-48149 Muenster, Germany}
\newcommand{\myongji}{Myongji University, Yongin, Kyonggido 449-728, Korea}
\newcommand{\nagasaki}{Nagasaki Institute of Applied Science, Nagasaki-shi, Nagasaki 851-0193, Japan}
\newcommand{\newmex}{University of New Mexico, Albuquerque, NM 87131, U.S. }
\newcommand{\nmsu}{New Mexico State University, Las Cruces, NM 88003, U.S.}
\newcommand{\ornl}{Oak Ridge National Laboratory, Oak Ridge, TN 37831, U.S.}
\newcommand{\orsay}{IPN-Orsay, Universite Paris Sud, CNRS-IN2P3, BP1, F-91406, Orsay, France}
\newcommand{\peking}{Peking University, Beijing, People's Republic of China}
\newcommand{\pnpi}{PNPI, Petersburg Nuclear Physics Institute, Gatchina, Leningrad region, 188300, Russia}
\newcommand{\riken}{RIKEN, The Institute of Physical and Chemical Research, Wako, Saitama 351-0198, Japan}
\newcommand{\rikjrbrc}{RIKEN BNL Research Center, Brookhaven National Laboratory, Upton, NY 11973-5000, U.S.}
\newcommand{\rikkyo}{Physics Department, Rikkyo University, 3-34-1 Nishi-Ikebukuro, Toshima, Tokyo 171-8501, Japan}
\newcommand{\saispbstu}{Saint Petersburg State Polytechnic University, St. Petersburg, Russia}
\newcommand{\saopaulo}{Universidade de S{\~a}o Paulo, Instituto de F\'{\i}sica, Caixa Postal 66318, S{\~a}o Paulo CEP05315-970, Brazil}
\newcommand{\seoulnat}{System Electronics Laboratory, Seoul National University, Seoul, South Korea}
\newcommand{\stonybrkc}{Chemistry Department, Stony Brook University, Stony Brook, SUNY, NY 11794-3400, U.S.}
\newcommand{\stonycrkp}{Department of Physics and Astronomy, Stony Brook University, SUNY, Stony Brook, NY 11794, U.S.}
\newcommand{\subatech}{SUBATECH (Ecole des Mines de Nantes, CNRS-IN2P3, Universit{\'e} de Nantes) BP 20722 - 44307, Nantes, France}
\newcommand{\tenn}{University of Tennessee, Knoxville, TN 37996, U.S.}
\newcommand{\titech}{Department of Physics, Tokyo Institute of Technology, Oh-okayama, Meguro, Tokyo 152-8551, Japan}
\newcommand{\tsukuba}{Institute of Physics, University of Tsukuba, Tsukuba, Ibaraki 305, Japan}
\newcommand{\vandy}{Vanderbilt University, Nashville, TN 37235, U.S.}
\newcommand{\waseda}{Waseda University, Advanced Research Institute for Science and Engineering, 17 Kikui-cho, Shinjuku-ku, Tokyo 162-0044, Japan}
\newcommand{\weizmann}{Weizmann Institute, Rehovot 76100, Israel}
\newcommand{\yonsei}{Yonsei University, IPAP, Seoul 120-749, Korea}
\affiliation{\abilene}
\affiliation{\banaras}
\affiliation{\bnl}
\affiliation{\caucr}
\affiliation{\charlesczech}
\affiliation{\ciae}
\affiliation{\cns}
\affiliation{\colorado}
\affiliation{\columbia}
\affiliation{\czechtech}
\affiliation{\dapnia}
\affiliation{\debrecen}
\affiliation{\elte}
\affiliation{\fit}
\affiliation{\fsu}
\affiliation{\gsu}
\affiliation{\hiroshima}
\affiliation{\ihepprot}
\affiliation{\illuiuc}
\affiliation{\instpasczech}
\affiliation{\isu}
\affiliation{\jinrdubna}
\affiliation{\kek}
\affiliation{\kfki}
\affiliation{\korea}
\affiliation{\kurchatov}
\affiliation{\kyoto}
\affiliation{\labllr}
\affiliation{\lawllnl}
\affiliation{\losalamos}
\affiliation{\lpc}
\affiliation{\lund}
\affiliation{\muenster}
\affiliation{\myongji}
\affiliation{\nagasaki}
\affiliation{\newmex}
\affiliation{\nmsu}
\affiliation{\ornl}
\affiliation{\orsay}
\affiliation{\peking}
\affiliation{\pnpi}
\affiliation{\riken}
\affiliation{\rikjrbrc}
\affiliation{\rikkyo}
\affiliation{\saispbstu}
\affiliation{\saopaulo}
\affiliation{\seoulnat}
\affiliation{\stonybrkc}
\affiliation{\stonycrkp}
\affiliation{\subatech}
\affiliation{\tenn}
\affiliation{\titech}
\affiliation{\tsukuba}
\affiliation{\vandy}
\affiliation{\waseda}
\affiliation{\weizmann}
\affiliation{\yonsei}
\author{A.~Adare}	\affiliation{\colorado}
\author{S.~Afanasiev}	\affiliation{\jinrdubna}
\author{C.~Aidala}	\affiliation{\columbia}
\author{N.N.~Ajitanand}	\affiliation{\stonybrkc}
\author{Y.~Akiba}	\affiliation{\riken} \affiliation{\rikjrbrc}
\author{H.~Al-Bataineh}	\affiliation{\nmsu}
\author{J.~Alexander}	\affiliation{\stonybrkc}
\author{K.~Aoki}	\affiliation{\kyoto} \affiliation{\riken}
\author{L.~Aphecetche}	\affiliation{\subatech}
\author{R.~Armendariz}	\affiliation{\nmsu}
% ...
\collaboration{PHENIX Collaboration} \noaffiliation

\date{\today}

\begin{abstract}
The PHENIX experiment presents results from the RHIC 2005 run with 
polarized proton collisions 
at $\sqrt{s}=200$ GeV, for inclusive $\pi^{0}$ 
production at mid-rapidity.  Unpolarized cross section results are  given
for transverse momenta $p_T=0.5$ to $20$ GeV/$c$, extending the range of 
published data to both lower and higher $p_T$.  
The cross section is described well for
$p_T<1$ GeV/$c$ by an exponential in $p_T$, and, for $p_T>2$ GeV/$c$, 
by perturbative QCD.  
Double helicity asymmetries $A_{LL}$ are presented based on a factor of five 
improvement in uncertainties as compared to previously published results, 
due to both an improved beam polarization of 50\%, and to higher integrated 
luminosity.  
These measurements are sensitive to the gluon polarization in the proton, 
and exclude maximal values for the gluon polarization.  

\end{abstract}

	
% For heavy ion papers we usually use just the one above (max is 4)
%%%%%%%%% Examples for p+p and spin papers include:
% PPG031:  \pacs{14.20.Dh, 13.60.Hb, 21.10.Hw, 25.40.Fq}
% PPG050:  \pacs{14.20.Dh, 25.40.Ep, 13.85.Ni, 13.88.+e}
% PPG037:  \pacs{13.85.Qk, 13.20.Fc, 13.20.He, 25.75.Dw}

% It is optional to also add (uncomment):
% \keywords{}

\maketitle

A principal goal of the spin program at the Relativistic Heavy Ion
Collider (RHIC) at Brookhaven National Laboratory is to determine 
the gluon spin contribution to a longitudinally polarized proton ($\Delta G$), 
taking advantage of the strongly 
interacting probes available in proton-proton collisions \cite{rhic_spin}. 
Previous measurements have established the validity of the perturbative
Quantum Chromodynamics (pQCD) description for inclusive mid-rapidity 
$\pi^{0}$ \cite{pi0cross_run2} and forward $\pi^{0}$ production~\cite{pi0_star}, 
and for mid-rapidity jet~\cite{jet_star} and direct photon 
production~\cite{photon_phenix}, at $\sqrt{s}=200$ GeV.
The double helicity asymmetries for the production of these particles 
involve gluons in the hard scattering processes in this pQCD 
description, and the first measurements for 
$\pi^{0}$ \cite{pi0all_run3,pi0all_run4} and for
jets \cite{jet_star} have begun to probe $\Delta G$.

The RHIC beam polarization and 
luminosity have significantly improved \cite{acc_performance}. 
The statistical uncertainty for a double helicity asymmetry
measurement is proportional to the inverse of $P^{2} \times \sqrt{{\cal L}}$ 
for beam polarizations $P$ and integrated luminosity ${\cal L}$, 
and decreased by a factor of 5 from the previously published data from 
PHENIX~\cite{pi0all_run3,pi0all_run4}.

In this paper, we first present the cross section for mid-rapidity $\pi^{0}$ 
production for unpolarized proton-proton collisions at $\sqrt{s}=200$ GeV.  
These results extend to lower and higher $p_T$ than in previous publications, 
and we discuss an apparent transition region between soft 
and hard scattering; the inclusive cross section is dominated by hard 
scattering, described by pQCD, for $p_T>2$ GeV/$c$. 
We then present the double helicity asymmetry, 
$A_{LL}$, for mid-rapidity $\pi^{0}$ production.
We also include measurements of $A_{LL}$ at low $p_{T}$, below the hard 
scattering region.  Finally, our results for $p_T>2$ GeV/$c$ are compared 
to a pQCD calculation that incorporates a model of gluon polarization.  
We present the range that we probe in 
the gluon momentum fraction ($x_{g}$) and discuss the constraint 
from these data on $\Delta G$.

The PHENIX experiment at RHIC measured $\pi^0$'s via 
$\pi^0 \rightarrow \gamma\gamma$ decays using a highly segmented 
($\Delta \eta \times \Delta \phi \sim 0.01 \times 0.01$) 
electromagnetic calorimeter (EMCal)~\cite{nim_emc}, covering a pseudorapidity 
range of $|\eta| < 0.35$ and azimuthal angle range of $\Delta \phi = \pi$. 
The $\pi^0$ data in this analysis were collected using two different 
trigger conditions. A minimum bias (MB) trigger was defined by the coincidence 
of signals in two beam-beam counters (BBC) with full azimuthal coverage 
located at pseudorapidities $\pm (3.0-3.9)$~\cite{nim_bbc}.
The cross section for events selected by the MB trigger was 
23.0~mb (about half of $\sigma^{inel}_{pp}$) with a systematic uncertainty 
of $\pm 9.7\%$ , derived from vernier 
scan results \cite{pi0cross_run2} and the variation of MB trigger efficiency 
for subsequent years. Higher $p_T$ data 
were collected using the coincidence of the MB trigger and an EMCal-based 
high $p_T$ photon trigger \cite{pi0cross_run2,pi0cross_run3}, 
with efficiency $\sim5\%$ at $p_T(\pi^0)\sim1$ GeV/$c$ and $\sim90\%$ 
for $p_T(\pi^0)>3.5$ GeV/$c$.  
The collision vertex was required to be within $|z|<30$ cm 
along the beam axis, based on the time difference between the 
two BBC detectors. The $\pi^0$ acceptance is uniform over this interval. 
The analyzed data sample of the 2005 run corresponds to an integrated 
luminosity of 2.5 pb$^{-1}$.

Details of the unpolarized cross section analysis technique are described in 
\cite{pi0cross_run2,pi0cross_run3}.
The background contribution under the $\pi^0$ peak in the two-photon invariant 
mass distribution varied from $~80\%$ in the lowest 0.5--0.75 GeV/$c$ 
$p_T$ bin to less than $8\%$ for $p_T>4$ GeV/$c$. 
The $\pi^0$ spectrum was corrected for overlapping decay photon showers in 
the EMCal, based on Monte Carlo simulations confirmed with test beam 
data~\cite{ieee_prec}. Below a $p_T(\pi^{0})$ of 12 GeV/$c$ the correction 
is less than $4\%$, and for $p_{T}(\pi^{0})=20$ GeV/$c$ the correction is 
$\sim$25\% and $\sim$70\%, for two different EMCal subsystems~\cite{nim_emc}. 
The systematic uncertainty of the measurement 
(excluding the $9.7\%$ uncertainty from the MB trigger cross section)
varied from $\sim 7\%$ at $p_T \sim 1$ GeV/$c$ to $\sim 16\%$ for the 
highest $p_T$ bin. 

%Measurements of $\pi^0$ cross section in p+p collisions at 
%$\sqrt{s}=200$ GeV were published in \cite{pi0cross_run2, pi0cross_dAu}.
%The analysed high $p_T$ data sample in this paper is about an order of magnitude larger than 
%published in \cite{pi0cross_dAu}, which allowed to considerably decrease the statitical errors 
%of the measurements and extend the measured $p_T$ range to $\sim$20 GeV/c. 
%Further, to enable a systematic study of the low $p_{T}$ end of the inclusive pion spectrum
%we for the first time include $\pi^{0}$ measurements with $p_{T}< 1$ GeV/c.

%Details of the analysis method are described in ~\cite{pi0all_run3}. 

\begin{figure}[tbh]
\includegraphics[width=1.0\linewidth]{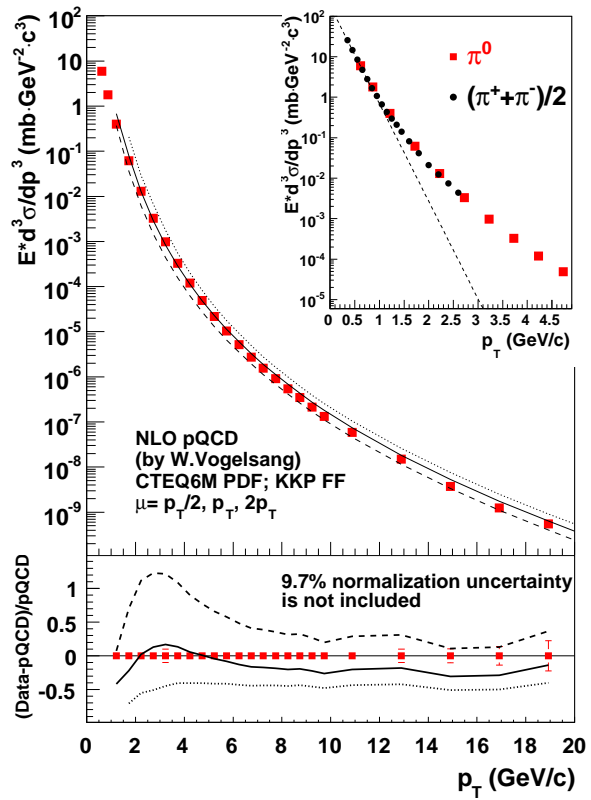}
\caption{\label{fig:cross} The neutral pion production cross section 
at $\sqrt{s}=200$ GeV as a function of $p_{T}$ (squares) 
and the results of NLO pQCD calculations for theory scales 
$\mu=p_T/2$ (dotted line), $p_T$ (solid line) and $2p_T$ (dashed line),
see text for details; 
note that the error bars are smaller than the points. 
The inset shows, in addition to $\pi^0$, data for $(\pi^{+}+\pi^{-})/2$ 
(solid circles), and a fit of charged pion data to an exponential function 
for $p_T<0.8$ GeV/$c$ (dashed line).
The bottom panel shows the relative difference between the data and 
theory for the three theory scales. 
Error bars are quadratic sums of experimental statistical and systematic 
uncertainties (the $9.7\%$ normalization uncertainty is not included). 
}
\end{figure}

Figure \ref{fig:cross} presents the cross section results for mid-rapidity $\pi^0$ 
production at $\sqrt{s}=200$ GeV, versus $p_T$, from $p_T$=0.5 GeV/$c$ to 
$p_T$=20 GeV~\cite{data}. 
Points are plotted at the average $p_T$ for each bin.  
The pQCD prediction, at next-to-leading order, is shown 
for theory scales $\mu=p_T/2$, $p_T$ and $2p_T$, where $\mu$ represents 
equal factorization, renormalization, and fragmentation scales~\cite{acgg,jsv}. 
The CTEQ6M parton distribution functions~\cite{cteq6m} 
and KKP set of fragmentation functions~\cite{frag_theory:kkp} are used. 
These data extend the published cross section data at both low and  
high $p_T$, and are consistent with previously published 
results~\cite{pi0cross_run2,pi0cross_run3}.
From $p_T$=2~GeV/$c$ to 20 GeV/$c$, the NLO pQCD calculation describes the data 
over a change in cross section of seven orders of magnitude.

The inset to Fig.~\ref{fig:cross} shows the lower $p_T$ region in more 
detail including 
high precision data for the charged pion cross section from \cite{phenix_ch}.
The data show a transition in the $p_T$ dependence of the cross section, from
exponential to a power law dependence, in the region $p_T\approx$1--2~GeV/$c$.  
In order to estimate possible contamination from non-perturbative physics
in the higher $p_T$ data, an exponential function ($\sim e^{-\alpha p_T}$) 
representing a non-perturbative component, is fit to the charged pion spectrum 
in the region $p_T$=0.3 to 0.8 GeV/$c$ (only the lowest $p_T$ $\pi^0$ 
data point is in this range) and extrapolated to the higher $p_T$ region. 
The exponential fit for the low $p_T$ region
gives $\alpha=5.56 \pm 0.02$~(GeV/$c$)$^{-1}$, 
with $\chi^2/NDF=6.2/3$. Only statistical uncertainties for the charged pion 
data were used in the fit. The dominant systematic uncertainty for the  
points in the fitted $p_T$ range is a $\sim12\%$ normalization uncertainty 
(excluding the normalization uncertainty from the MB trigger cross section).
Beyond about $p_T$=1~GeV/$c$, the data lie above this single exponential.  
The fraction of the exponential contribution to the data for the
2--2.5 GeV/$c$ $p_T$ bin is found to be less than 10\%, 
with a negligible contribution for higher $p_T$.  
This is the basis for applying the pQCD formalism
to the double helicity asymmetry data with $p_T>2$ GeV/$c$.

For the 2005 run, each collider ring of RHIC was filled with up to 
111 bunches in a 120 bunch pattern, spaced 106 ns apart, with predetermined 
patterns of polarization signs for the bunches. 
Spin rotators, sets of four helical dipole magnets 
on each side of PHENIX, rotate the polarization orientation from vertical, 
the stable spin direction in the RHIC arcs, to longitudinal at the interaction 
point~\cite{waldo}.
Beam helicity asymmetries are obtained by tagging the polarization signs of the
bunches for each event.  The bunches for one beam alternate in polarization 
sign, and pairs of bunches alternate in sign for the other beam.  
In this way data for all combinations of beam helicity are collected at 
the same time, and the possibility of false asymmetries due to changing 
detector response versus spin state are greatly reduced. 
Each RHIC fill, typically lasting 8 hours, used one of four bunch 
spin patterns.

The beam polarizations for 2005 were measured using fast carbon 
target polarimeters~\cite{pol_pC}, normalized by absolute polarization 
measurements made during 2005 by a separate polarized atomic hydrogen jet 
polarimeter~\cite{pol_jet}. 
The beam polarizations, from luminosity-weighted averages over 104 RHIC 
fills used in the analysis, were 
$\langle P^B \rangle$=0.50$\pm$0.002(stat)$\pm$0.025(systB)$\pm$0.015(systG)
and
$\langle P^Y \rangle$=0.49$\pm$0.002(stat)$\pm$0.025(systY)$\pm$0.015(systG),
for ``Blue'' and ``Yellow'' RHIC beams, respectively, 
for the bunches colliding at PHENIX.  
The systematic uncertainties have been separated into uncorrelated 
uncertainties for each beam, ``systB'' and ``systY'', 
and a global systematic uncertainty ``systG'', 
which is common for both beams and comes from systematic uncertainty in 
jet polarimeter measurements~\cite{pol_2005}.
%The uncorrelated systematic uncertainties include uncertainties 
%due to observed and possible variations of polarization over the
%beam transverse profiles, and the statistical uncertainties of the 
%jet target calibrations of each carbon target polarimeter. [ref]  
%The global uncertainty includes uncertainties from
%background in the jet target calibration and in the unpolarized 
%molecular fraction of the polarized atomic hydrogen jet target. [ref] 
For comparison, the polarizations in the 2004 run were 0.44$\pm$0.08(syst).

Local polarimeters based on very forward neutron 
production (production angle 0.3--2.5 mrad)~\cite{locpol,pi0all_run3} 
were used to set up and monitor 
the beam polarization orientation at PHENIX.  
The polarimeters monitor the transverse
polarization of each beam at PHENIX, which can be compared to the beam 
polarization measured by the RHIC polarimeters where the polarization 
direction is vertical.
The local polarimeters were calibrated by turning off the spin rotators around 
PHENIX, and measuring the response of the local polarimeters with the beams 
vertically polarized.  
%The local polarimeters are nearly symmetric in azimuth and in general measure 
%the degree of transverse polarization at PHENIX. 
For the longitudinal polarization data, the beams showed a measurable
transverse polarization, with $(P_T/P)^B$=0.10$\pm$0.02 
and $(P_T/P)^Y=0.14\pm0.02$,
with $P_T/P$ referring to the fraction of transverse polarization of each beam.  
The polarization directions, as determined by the spin rotator settings 
and as measured by 
the local polarimeters, remained constant over the run. The product of the 
beam polarizations $P^B \cdot P^Y$ is required for the double helicity asymmetry 
measurement. The average transverse component of the product was 
$\langle P_T^B \cdot P_T^Y \rangle / \langle P^B \cdot P^Y \rangle < (P_T/P)^B \cdot (P_T/P)^Y = 0.014\pm0.003$; the average of the  polarization 
product over the run was $\langle P^B \cdot P^Y \rangle = 0.24$ 
with a systematic uncertainty of $\pm9.4\%$.

The double helicity asymmetry $A_{LL}$ is the difference of cross sections for 
the same versus opposite beam helicities, divided by the sum.  
Experimentally, for inclusive $\pi^0$ production, it can be determined as: 
\begin{equation}
A_{LL}^{\pi^0} = \frac{1}{|P^B \cdot P^Y|} \cdot \frac{N_{++}-R\cdot
N_{+-}}{N_{++} + R \cdot N_{+-}};~~~R=\frac{L_{++}}{L_{+-}},
\label{eq:a_ll}
\end{equation}
where $N$ is the number of $\pi^0$'s measured in PHENIX from the colliding bunches 
with the same ($++$) and opposite ($+-$) helicities, and $R$ is the relative 
luminosity between bunches with the same and opposite helicities.
Here we neglect the parity-violating difference in cross section between 
$(++) \leftrightarrow (--)$ and $(+-)  \leftrightarrow (-+)$ beam helicity 
configurations~\cite{a_l}.
$A_{LL}$ was calculated for each fill in order to reduce systematics from 
variation in beam polarizations and in $R$ for different fills. 
Even and odd crossings were handled by separate high $p_T$ photon trigger 
electronics chains. To avoid possible detector bias, 
$A_{LL}$ was also determined separately for the even and odd crossings. 
Final asymmetries were averaged, and corrected 
for the asymmetry of the background under the $\pi^0$ peak 
in the two-photon mass distribution, as in~\cite{pi0all_run3}. 

The relative luminosity ratio $R$ is obtained from the minimum bias triggers (MB) 
discussed above. Scalers keep track of the number of live triggers for each 
bunch crossing. Single beam background was $<0.05\%$,
as measured from non-colliding bunches, and contributes negligible systematic 
uncertainty to the measured $R$.
We also measured the double helicity asymmetry of the relative luminosity 
scaler counts, by normalizing using zero degree neutral particle production 
as measured by zero degree calorimeters (ZDC)~\cite{nim_zdc}. 
No asymmetry was observed. This gave a limit on an asymmetry bias in the 
measurement of $\delta A^{\pi^0}_{LL}|_{\rm bias}<2\times10^{-4}$, and a limit on the 
systematic uncertainty for the measurement of relative luminosity 
giving $\delta A^{\pi^0}_{LL}|_{R}<2\times10^{-4}$.
These limits also include the effects from the pileup of two 
or more collisions in a crossing, calculated at  $ \alt 4\%$ of the crossings.
The BBC and ZDC monitors observe the pileup at significantly different rates, 
and therefore the limits above, from comparing BBC and ZDC counts, 
include these uncertainties. 

A transverse double spin asymmetry $A_{TT}$, the transverse equivalent 
to Eq.~(\ref{eq:a_ll}), can contribute to $A_{LL}$ through the 1.4\% 
transverse component of the product of the beam polarizations discussed above. 
Although $A_{TT}$ has been postulated to be extremely small, 
$\sim 10^{-4}$~\cite{att}, it has not been previously measured.
We measured $A_{TT}$ in a short run with transverse polarization. 
$A_{TT}(p_T)$ was consistent with zero within statistical errors~\cite{data};
the errors were 5 times larger than the uncertainties for $A_{LL}$, 
$\delta^{stat} A_{LL}$. Therefore, a limit was determined for the 
$A_{TT}$ contribution to $A_{LL}$ of $0.07 \cdot \delta^{stat} A_{LL}$.

\begin{figure}[tbh]
	\includegraphics[width=1.0\linewidth]{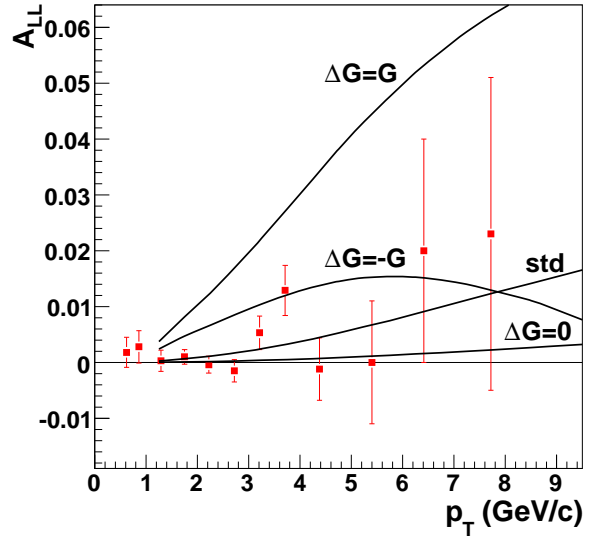}
	\caption{\label{fig:all} The double helicity asymmetry for 
	  neutral pion production at $\sqrt{s}=200$ GeV 
	  as a function of $p_{T}$ (GeV/$c$). 
	  Error bars are statistical uncertainties, 
	  with the 9.4\% scale uncertainty not shown; 
	  other experimental systematic uncertainties are negligible. 
	  Four GRSV theoretical calculations based on NLO pQCD are also shown for 
	  comparison with the data (see text for details.)
	}
\end{figure}

Figure \ref{fig:all} presents the measured double helicity asymmetry in  
$\pi^{0}$ production~\cite{data}. 
A scale uncertainty of 9.4\% in $A_{LL}^{\pi^0}$
due to the uncertainty in beam polarization is 
not shown. The other systematic uncertainties are negligible, as 
discussed above, and checked using a bunch polarization sign randomization 
technique, and by varying the $\pi^0$ identification criteria~\cite{pi0all_run3}. 
Data for $p_T>1$ GeV/$c$ were obtained from the high $p_T$ photon 
triggered sample. For $p_T$ below 1 GeV/$c$, due to low efficiency for the 
high $p_T$ photon trigger, we used the MB data sample. 
In the low $p_T$ region, where the cross section shows an exponential behavior, 
the helicity asymmetry is $A_{LL}^{\pi^0}$=0.002$\pm$0.002,
for the data in the range $p_T=0.5-1$~GeV/$c$. 
For the higher $p_T$ region, the four curves in Fig.~\ref{fig:all} show calculations 
of $A_{LL}^{\pi^0}$, using NLO pQCD with $\mu=p_T(\pi^0)$, that reflect 
the range of gluon polarizations allowed by inclusive deep inelastic 
scattering (DIS) data. 
The calculations are based on the GRSV model, where ``std'' was the best fit 
to inclusive DIS data~\cite{grsv}. 
For momentum fraction $x$, $\Delta G(x)=G^{+}(x)-G^{-}(x)$ refers to the 
gluon helicity distribution, and $G^{+}(x)$ and $G^{-}(x)$ refer to the 
gluon densities for $+$ and $-$ helicities in a $+$ helicity proton.
The first moment of the gluon helicity distribution, 
$\int \limits_{0}^{1} \Delta G(x) dx$, for the ``std'' parameterization
is $\Delta G$=0.4, at the scale $Q^2$=1~GeV$^2$. 
The other three curves are calculations based on this 
best fit, but use at $Q^{2}=0.4$ GeV$^{2}$ the function 
$\Delta G(x) = G(x), 0, -G(x)$, where $G(x)$ is the unpolarized gluon 
distribution. The gluon distribution at the input scale is evolved to the 
scale $Q^2=p_{T}^{2}(\pi^0)$.

%To estimate the effect of the beam residual transverse 
%polarization to $A_{LL}$, we extracted $A_{TT}$ from the $\pi^0$ data sample 
%obtained from the vertically polarized proton collisions. Double transverse 
%spin asymmetries were found to be consistent with zero in all $p_T$ bins 
%within statistical errors $\delta^{stat} A_{TT}$, which were larger 
%by a factor about 5, than $\delta^{stat} A_{LL}$, due to smaller statistics 
%collected in transversally polarized data sample. Taking into account 
%the small residual transverse spin contribution in the longitudinally 
%polarized data sample, the possible effect of $A_{TT}$ to $A_{LL}$ was 
%found to be less than 10\% of $\delta^{stat} A_{LL}$.

In order to explore the impact of the new data on the
sensitivity to the polarized gluon distribution, 
we have compared the data with a set of $A_{LL}(p_{T})$
curves corresponding to different $\Delta G(x)$ between 
$\Delta G(x) = -G(x)$ and $\Delta G (x) =G(x)$ at $Q^{2}=0.4$ GeV$^{2}$.
We used the data for $p_T>2$~GeV/$c$, 
which appear to have little contamination from soft physics as discussed earlier. 
The most likely $x_g$ for PHENIX $\pi^0$ data in each $p_T$ point is 
$\sim x_T/0.8$ \cite{rak}, where $x_T=p_T/(\sqrt{s}/2)$. 
For the measured $p_T$ range 2--9 GeV/$c$, the range of $x_g$ in each bin is broad, 
and spans the range $x_g=0.02-0.3$, as calculated by NLO pQCD~\cite{werner_private}.

\begin{figure}[tbh]
\includegraphics[width=1.0\linewidth]{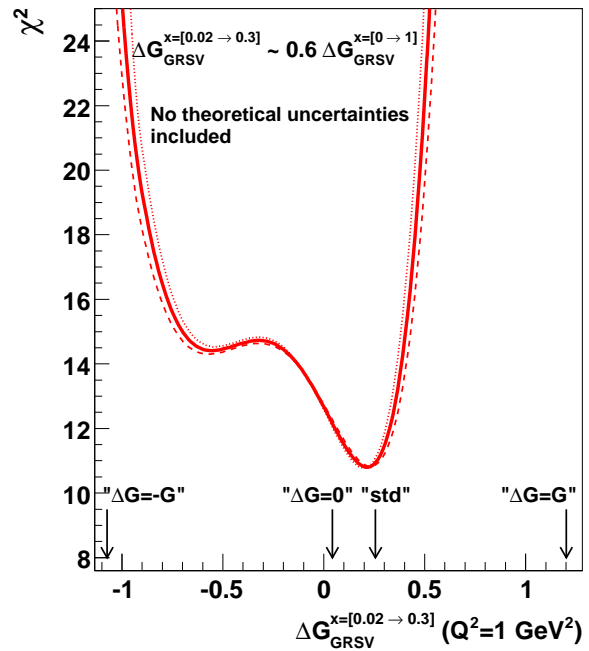}
\caption{\label{fig:chi2} 
The $\chi^{2}$ distribution of the measured data
plotted versus the value of the first moment of the polarized gluon distribution 
(solid line) in the $x_g$ range from 0.02 to 0.3 
corresponding to our $\pi^0$ data in $p_{T}$ bins from 2 to 9 GeV/$c$.
Dashed and dotted lines correspond to $-9.4\%$ and $+9.4\%$ variation in 
$A_{LL}$ normalization related to the beam polarization uncertainty, the 
dominant systematical uncertainty of our data. 
Only statistical uncertainties were used for each curve.
Arrows indicate $\Delta G$ corresponding to the different polarized gluon 
distributions discussed in the text.
%The $\Delta G$ range on the plot corresponds to the $\Delta G = \pm G$ 
%curves on Figure \ref{fig:all}.
}
\end{figure}

Figure~\ref{fig:chi2} shows the corresponding $\chi^{2}$ versus 
$\Delta G_{GRSV}^{x=[0.02 \rightarrow 0.3]}$, where we compare to an integral 
of $\Delta G$ over the probed $x_g$ range.
Only experimental statistical uncertainties are used to calculate $\chi^2$, 
and no theoretical uncertainties are included. 
It is important to note, that although the range of the first
moment explored represents $\sim60\%$ of the full integral, 
this reflects using a specific model for the gluon polarization.
For example, a gluon polarization model with a crossover from positive to negative 
gluon polarization within our $x_{g}$ range would yield a small average asymmetry 
for each point. Also, other models can generate larger or smaller contributions 
from the gluon spin in the unmeasured region of $x_{g}$.

These data are sensitive to the first moment of the polarized gluon
distribution.  
Using the GRSV model, we find that the gluon polarization contribution to the proton
spin (1/2) in the probed $x_{g}$ range is constrained between $-$0.9 and $+$0.5, 
for $\chi^2-\chi^2_{min}$=9, representing a ``3$\sigma$'' limit 
(a ``1$\sigma$'' limit would give a constraint between 0.07 and 0.3).
The extremes of gluon 
polarization are ruled out, modulo the above remarks,
with the confidence level for ``$\Delta G =\pm G$'' of less than $10^{-6}$. 
Large positive gluon polarization~\cite{deltaG_large} was proposed shortly after 
the discovery that the quark contribution to the proton spin was small~\cite{EMC}, 
with the suggestion that such a large gluon polarization
would mask a ``bare'' quark polarization. 
For ``std'' and ``$\Delta G=0$'', the confidence levels are 20--21\% and 12--13\%,
respectively, for the range of $\pm9.4\%$ scale uncertainty of the measurement.  
Semi-inclusive DIS measurements ~\cite{sDIS} have also presented data on 
$\Delta G$ in a limited $x_g$ range and its comparison with various 
$\Delta G$ scenarios. 

The two minima in Fig.~\ref{fig:chi2} reflect the quadratic contribution of
the gluon polarization to $A_{LL}$, from the gluon-gluon scattering subprocess 
for $\pi^0$ production. The symmetry between the two minima is broken by 
the quark-gluon scattering subprocess, where the gluon polarization contributes 
linearly to $A_{LL}$. The quark-gluon subprocess is emphasized at higher $p_T$, 
which will become accessible with additional running at high polarization 
and luminosity. 

%Compared to our previous result, we have made a significantly better 
%measurement of the $A_{LL}^{\pi^{0}}$: 20 times better in terms of 
%the polarization and luminosity related figure of merit; and which is  
%consistent with our previous measurements.  
%The extraction of $\Delta G$ from the world set of DIS and these new
%RHIC data, requires a global next to leading order pQCD analysis. 
%We have explored the potential impact of the $\pi^{0}$ data 
%for constraining the  contribution of the gluons to the proton spin 
%using representative QCD analysis  [GRSV].
%We demonstrated that addition of our data to the existing QCD  analyses of
%the pDIS data, further constraints the $\Delta G(x)$ in our probed
%kinematic region. Larger data set of $\pi^{0}$, particularly at high  $p_{T}$,
%along with other probes of the gluon distribution such as charged pions,
%prompt photons will help to resolve the sign ambiguity of this constrain. 

To summarize, we have presented the unpolarized cross section and double 
helicity asymmetries for $\pi^{0}$ production at mid-rapidity, 
for proton-proton collisions at $\sqrt{s}=200$ GeV.  We observe an apparent 
transition region in the cross section, for $p_{T} \approx$1 to 2 GeV/$c$, 
with the cross section described by an exponential in $p_{T}$ 
below $p_{T} \sim 1$ GeV/$c$, and with the cross section described  
by the pQCD prediction for $p_{T}=2$ to $20$ GeV/$c$, over seven orders 
of magnitude in cross section. The results
for $A_{LL}$ in the pQCD region, which we take as $p_{T } \geq 2$ GeV/$c$, 
constrain the gluon polarization in the proton significantly.  
The range probed is $x_{g} = 0.02$ to $0.3$, for the gluon momentum 
fraction.  Using one representative model for the gluon polarization, 
GRSV \cite{grsv}, which assumes no crossover in gluon polarization 
versus $x_{g}$, we present a map of $\chi^{2}$ versus the first 
moment of the polarized gluon distribution in the measured region.  
%Using $\chi^{2} - \chi^{2}_{\rm min} < 9 $ ($3\sigma$), we obtain 
%$\Delta G_{GRSV}^{x=0.02-0.3} = -0.5~to +0.3$, compared to the proton 
%spin of 1/2. 
%No theory uncertainties, such as different models 
%or theory scales, have been included.  
From this study, the present data 
rule out extreme values of gluon polarization suggested after the surprise of 
the EMC result that the quarks (and anti-quarks) contribute little to the 
spin of the proton \cite{EMC}, but allow significant contribution 
from the gluon spin to the proton spin. 

% See NOTES below for information about reference citations, figures 
% and tables, using wide text for equations, landscape figures, etc.

%%%%%%%%%%%%%%%%%%%%%%%%%  Acknowledgements 

\begin{acknowledgments}

We thank the RHIC Polarimeter Group and
the staff of the Collider-Accelerator and
Physics Departments at BNL for their vital contributions.
We thank W.~Vogelsang for providing the NLO pQCD calculations 
and M.~Stratmann for informative discussions. 
We acknowledge support from 
the Department of Energy and NSF (U.S.A.), 
MEXT and JSPS (Japan), 
CNPq and FAPESP (Brazil), 
NSFC (China), 
MSMT (Czech Republic),
IN2P3/CNRS, and CEA (France), 
BMBF, DAAD, and AvH (Germany), 
OTKA (Hungary), 
DAE (India), 
ISF (Israel), 
KRF and KOSEF (Korea), 
MES, RAS, and FAAE (Russia),
VR and KAW (Sweden), 
U.S. CRDF for the FSU, 
US-Hungarian NSF-OTKA-MTA, 
and US-Israel BSF.

\end{acknowledgments}

%%%%%%%%%%%%%%%%%%%%%%%%%%%  References 

\def\PRL{Phys. Rev. Lett.\ }
\def\PRD{{Phys. Rev.}~{\bf D}}
\def\PRC{{Phys. Rev.}~{\bf C}}
\def\PLB{{Phys. Lett.}~{\bf B}}
\def\NIMA{{Nucl. Instrum. Methods}~{\bf A}}


\begin{references}

\bibitem{rhic_spin} G.~Bunce {\it et al.}, Ann. Rev. Nucl. Part. Sci. {\bf 50}, 
525 (2000).

\bibitem{pi0cross_run2} S.S.~Adler {\it et al.}, \PRL {\bf 91}, 241803 (2003).

\bibitem{pi0_star} J.~Adams {\it et al.}, \PRL {\bf 92}, 171801 (2004).

\bibitem{jet_star} B.I.~Abelev {\it et al.}, \PRL {\bf 97}, 252001 (2006).

\bibitem{photon_phenix} S.S.~Adler {\it et al.}, \PRL {\bf 98}, 012002 (2007).

\bibitem{pi0all_run3} S.S.~Adler {\it et al.}, \PRL {\bf 93}, 202002 (2004).

\bibitem{pi0all_run4} S.S.~Adler {\it et al.}, \PRD {\bf 73}, 091102 (2006).

\bibitem{acc_performance} M.~Bai {\it et al.}, Proceedings of 
the 2005 Particle Accelerator Conference, edited by C.Horak, IEEE, p.600.

\bibitem{nim_emc}L.~Aphecetche {\it et al.}, \NIMA {\bf 499}, 521 (2003).

\bibitem{nim_bbc} M.~Allen {\it et al.}, \NIMA {\bf 499}, 549 (2003).

\bibitem{pi0cross_run3} S.S.~Adler {\it et al.}, \PRL {\bf 98},172302 (2007).

\bibitem{ieee_prec} G.~David {\it et al.}, IEEE Trans.Nucl.Sci.47: 1982-1986, 
2000.

\bibitem{data} Tables of data available at http://www.phenix.bnl.gov
/phenix/WWW/info/data/ppg063\_data.html

\bibitem{acgg} F.\ Aversa {\it et al.}, Nucl. Phys. {\bf B327}, 105 (1989); {\it n.b.}, 
these authors wrote the computer code used for our calculations.

%\bibitem{jsv}
%B.\ J\"{a}ger, A.\ Sch\"{a}fer, M.\ Stratmann, and W.\ Vogelsang,
%Phys. Rev. D {\bf 67}, 054005 (2003).

\bibitem{jsv} B.\ J\"{a}ger {\it et al.}, \PRD{\bf 67}, 054005 (2003).

\bibitem{cteq6m} J.~Pumplin {\it et al.}, 
%[CTEQ Collaboration], 
J. High Energy Phys. {\bf 07}, 012 (2002).

\bibitem{frag_theory:kkp} B.A.~Kniehl {\it et al.}, Nucl. Phys. {\bf B597}, 337 (2001).
%B.A.~Kniehl, G.~Kramer, and B.~P\"{o}tter, 

\bibitem{phenix_ch} S.S.~Adler {\it et al.}, \PRC {\bf 74}, 024904 (2006).

\bibitem{waldo} W.W.~MacKay {\it et al.}, Proceedings of 
the 2003 Particle Accelerator Conference, 
edited by J.~Chew, P.~Lucas and S.~Webber, IEEE, p.1697.

\bibitem{pol_pC} O.~Jinnouchi {\it et al.}, 
RHIC/CAD Accelerator Physics Note 171 (2004).

\bibitem{pol_jet} H.~Okada {\it et al.}, \PLB {\bf 638}, 450 (2006).

\bibitem{pol_2005} 
I.~Nakagawa {\it et al.}, RHIC/CAD Accelerator Physics Note 275 (2007); 
O.~Eyser {\it et al.}, RHIC/CAD Accelerator Physics Note 274 (2007).

\bibitem{locpol} Y.~Fukao {\it et al.}, hep-ex/0610030; submitted to \PLB.

\bibitem{a_l} PHENIX has measured parity-violating single helicity 
asymmetry $A_L$ for each polarized beam, which was consistent with zero 
within statistical uncertainty, in all $p_T$ bins \cite{pi0all_run3,data}.

\bibitem{nim_zdc} C.~Adler {\it et al.}, \NIMA {\bf 470}, 488 (2001).

\bibitem{att} A.~Mukherjee {\it et al.}, \PRD {\bf 72}, 034011 (2005).

\bibitem{grsv} B.~J\"ager {\it et al.}, \PRD {\bf 67}, 054005 (2003);
M.~Gl\"uck {\it et al.}, \PRD {\bf 63}, 094005 (2001).

\bibitem{deltaG_large} G. Altarelli, G. Ross, Phys. Lett. {\bf B212}, 391 (1988); 
G. Altarelli, W.  J. Stirling, Part. World {\bf 1}, 40; 
R. D. Carlitz, J. C. Collins, A. H. Mueller, Phys. Lett. {\bf B214}, 229 (1988).

\bibitem{EMC} EMC, J. Ashman {\em et al.}, Phys. Lett. {\bf B206} 
                   364 (1988), Nucl. Phys. {\bf B328}, 1 (1989).

\bibitem{rak} S.S.~Adler {\it et al.}, \PRD {\bf 74}, 072002 (2006).

\bibitem{werner_private} M.~Stratmann and W.~Vogelsang, hep-ph/0702083; 
W.~Vogelsang, private communication.

\bibitem{sDIS}
B.~Adeva {\it et al.} (SMC), \PRD {\bf 70}, 012002 (2004);
E.S.~Ageev {\it et al.} (COMPASS), Phys. Lett. {\bf B633}, 25 (2006); 
A.~Airapetian {\em et al.} (HERMES), Phys. Rev. Lett. {\bf 84}, 2584 (2000).

\end{references}
\end{document}